# Effect of Pb addition on microstructure, transport properties and the critical current density in a polycrystalline FeSe$_{0.5}$Te$_{0.5}$

Shiv J. Singh[1*], Ryszard Diduszko[2,3], Przemysław Iwanowski[3], Tomasz Cetner[1], Andrzej Wisniewski[3], Andrzej Morawski[1]

[1]*Institute of High Pressure Physics (IHPP), Polish Academy of Sciences, Sokolowska 29/37, Warsaw, Poland*

[2]*Lukasiewicz Research Network -Institute of Microelectronics and Photonics, al. Lotników 32/46, 02-668 Warsaw, Poland*

[3]*Institute of Physics, Polish Academy of Sciences, aleja Lotników 32/46, 02-668 Warsaw, Poland*

*Corresponding author:

Email: sjs@unipress.waw.pl

# Abstract

We have investigated the effects of lead (Pb) additions ($x$) up to 40 wt.% ($x$ = 0-0.4) on the structure, electrical properties, and magnetic properties of FeSe$_{0.5}$Te$_{0.5}$ superconductor. The samples were prepared by the solid-state reaction method and were characterized by various techniques. The parent compound ($x$ = 0) showed the onset temperature $T_c^{onset}$ of 15 K, and zero-resistance temperature, $T_c^{offset}$ of 12 K. The addition of Pb enhances the metallic characteristics of FeSe$_{0.5}$Te$_{0.5}$, but both $T_c^{onset}$ and $T_c^{offset}$ are decreased to the lower temperature with the broadened transition width. The $T_c^{onset}$ is nearly the same (10.3 K) at higher additions, such as $x$ = 0.3 and 0.4, but zero resistivity is not observed up to 7 K. Microstructural analysis and transport studies suggest that for $x$ > 0.05, Pb additions weakened the coupling between grains and suppressed the superconducting percolation, leading to a broad transition. More importantly, the inclusion of a relatively small amount of Pb ($x$ = 0.05) increased the critical current density, $J_c$, in the entire magnetic field, which might be attributable to better phase uniformity as well as good grain connectivity.

# Introduction

The iron-based superconductor (IBS) was discovered in 2008 [1], and following the pioneering work, many compounds have been discovered under this high $T_c$-family. More than 100 compounds are accessible as IBS today, which can be categorized into 6-7 families based on the crystal structure of their parent compounds [2, 3, 4, 5] such as oxypnictide *RE*OFeAs (1111) (*RE* = rare earth), $A$Fe$_2$As$_2$ (*A* = Ba, K, Ca) (122), FeSe$_x$Te$_{1-x}$ (11), CaKFe$_4$As$_4$ (1144), and LiFeAs (111). FeSe (11 family) has very simple crystal structures in all IBS families and exhibits a superconducting transition temperature ($T_c$) of 8 K in $\beta$-Fe$_x$Se [6, 7] at ambient pressure. At room temperature, the crystal lattice of $\beta$-Fe$_x$Se has a tetragonal (*P4/nmm*) symmetry and is made up of a layer-by-layer stack of edge-sharing FeSe$_4$-tetrahedra [7]. The planar sublattice of the layered Fe-based quaternary oxypnictides is identical to that of the tetragonal phase $\beta$-Fe$_x$Se with a PbO-type structure, which is the first system in which the Fe-element plays an important role in the occurrence of superconductivity. Following the initial work of the FeSe discovery [7], it's logical to question if chemical substitutions to the Se-site or Fe-site affect the superconducting properties of the material. As a result, several reports have been published based on various types of doping, such as Cu [8, 9], Ni [10], S [11, 12], Te [11], Cr [13], Co [11], either at Fe-sites or Se-sites in the FeSe system. Interestingly, the highest transition temperature, $T_c$, has been enhanced up to 37.6 K under the applied external pressure [6].

The 11-family is also free of rare earth elements and hazardous components. As a result, FeSe is attractive for a variety of applications, including superconducting tapes, wires, and superconducting magnets [14]. On the other hand, preparing single-phase superconducting material is difficult for this family. Hexagonal $\delta$-Fe$_x$Se, orthorhombic FeSe$_2$, tetragonal $\beta$-Fe$_x$Se, monoclinic Fe$_3$Se$_4$, and hexagonal Fe$_7$Se$_8$ [15] are some of the stable crystalline forms of iron selenides. The tetragonal $\beta$-Fe$_x$Se with a PbO structure (space group *P4/nmm*) transforms to an orthorhombic space group *Cmma* which exhibits a superconducting transition at ~8 K [7]. However, multiple phases generally appear in superconducting polycrystalline $\beta$-Fe$_x$Se samples [16] due to the complicated phase diagram of FeSe [15]. Some ferromagnetic contribution is also presented due to the remaining phase $\delta$-Fe$_x$Se (space group *P63/mmc*, NiAs-type), even if the extra phases $\alpha$-Fe and Fe$_7$Se$_8$ can be avoided by properly selecting the initial composition [16]. The $\beta$-Fe$_x$Se phase is non-magnetic, phase Fe$_7$Se$_8$ is antiferromagnetic, and especially, the unreacted $\alpha$-iron is ferromagnetic [16]. As a result, the magnetic contributions normally influence the superconducting properties of the 11-family. Although the magnetic contribution can be reduced by modifying the Se-content in the initial precursor material and the synthesis temperature setup, a fully non-magnetic sample cannot be obtained due to the presence of needle-like $\delta$-Fe$_x$Se phase that does not convert completely to $\beta$-Fe$_x$Se during the heat treatment procedure. Furthermore, we must note that the remaining high-temperature phase $\delta$-Fe$_x$Se, the unreacted $\alpha$-iron, Fe$_7$Se$_8$ phase presented in the samples, generates a magnetic background in the magnetization measurements that cannot be avoided. A recent study [16] shows that the Fe$_7$Se$_8$ phase appears in the final product when there is a high concentration of Se in the initial precursors, whereas unreacted iron is observed when Se is lost during the reaction process. Sometimes, these phases are difficult to identify through standard XRD measurements [16].

When FeSe with a tetragonal PbO-type structure is synthesized at roughly 460°C, it is claimed to undergo a phase change toward a hexagonal-NiAs structure. FeTe, on the other hand, with the same tetragonal crystal structure, is stable up to 930°C. The substitution of Se atoms within FeSe with Te would subsequently be expected to stabilize the tetragonal phase at a synthesis temperature at or above 460°C [17]. It is found that the superconducting transition temperature increases up to 15 K with Te doping, reaches a maximum at about 50% substitution, and then decreases with more Te doping [11, 18]. Interestingly, FeTe is no longer superconducting. At roughly 110 K, the parent compound FeSe undergoes a structural transformation from tetragonal (P4/*nmm*) to triclinic (*P*-1) symmetry [17], which alters the lattice properties without breaking magnetic symmetry and is thought to have a strong association with the development of superconductivity. This Te-substitution study confirms that structural deformation at low-temperatures is critical to the competition between magnetism and superconductivity in this class of materials.

Chemical addition is commonly used to improve superconducting properties by increasing superconducting phase crystallization, accelerating superconducting grain intergranular interaction, or introducing pinning centers [19, 4, 20, 21]. Carbon addition, for example, can considerably increase the irreversibility field and $J_c$ in MgB$_2$ [22], and a $J_c$ augmentation has been reported in Ag-added YBa$_2$Cu$_3$O$y$ (YBCO) [23]. Another important advancement in the development of the 122 pnictide (Sr$_{0.6}$K$_{0.4}$Fe$_2$As$_2$) wires was the addition of elements such as Ag or Pb to strengthen the intergranular coupling of the superconducting grains, resulting in the improved $J_c$ characteristics [19]. To improve superconducting properties and better understand the superconducting mechanism, various metal additions are also being studied for 11 family. Effects of several dopants such as Ag [24], Co [25], Ni [25], Li [26], Sn [27] have been reported to FeSe$_{0.5}$Te$_{0.5}$ bulks, where Li and Sn additions work very effectively to improve the superconducting properties. Although the metal Pb is a poisonous and volatile element, but an enhancement of $J_c$ in pnictide bulks and tapes was achieved by Pb addition [19]. However, there is no study available based on the Pb added FeSe$_{0.5}$Te$_{0.5}$ sample. High-pressure studies based on Ag added FeSe revealed the improved grain connectivity as well as the superconducting properties [28]. As discussed above, the tetragonal phase exhibits superconductivity in the 11 family, whereas the hexagonal phase is harmful to superconductivity and is inescapable in FeSe$_{0.5}$Te$_{0.5}$ sintered samples unless a long sintering period at around 410°C is used [29].

In this paper, we report a detailed study of the structure, microstructure, and the superconducting properties of FeSe$_{0.5}$Te$_{0.5}$ under the influence of Pb additions ($x$ = 0-0.4). These samples were characterized by X-ray powder diffraction (XRD), microstructure, resistivity, and magnetic measurements. The addition of Pb improved the metallic behavior of the parent compound. The microstructural analysis revealed that for $x > 0.05$, there was no improvement in the grain connections and enhancement of impurity phases with Pb additions. It is found that the transition temperature ($T_c$) is reduced with the Pb-additions, which could be due to the reduced concentration of Te in the FeSe$_{0.5}$Te$_{0.5}$ structure. Besides, a small amount of Pb addition can improve the critical current properties.

## Experimental details

We have prepared polycrystalline FeSe$_{0.5}$Te$_{0.5}$ samples by using the solid-state reaction method. In the first step, the purity of the starting reagents, *i.e.*, Fe powder (>99.5% purity, Alfa Aesar), Se (99.99% purity, Alfa Aesar), and Te (99.99% purity, Alfa Aesar) were mixed according to the stoichiometric ratios of FeSe$_{0.5}$Te$_{0.5}$, which were well-mixed in an agate mortar for 15-20 min. Then, these well-mixed powders were cold-pressed into discs under 6000 Kg of uniaxial pressure and sealed into an evacuated quartz tube with a pressure of less than 1.3 Pascal. The evacuated quartz tube is used to minimize the oxygen atmosphere inside the tube to the precursors and also helps to seal the tube. The prepared ampoules were heated at 600°C for 11 h in a box furnace. In one batch, we prepared around 14 grams of FeSe$_{0.5}$Te$_{0.5}$ sample, and the resultant powder was reground into a fine powder. To prepare 2 g sample for each Pb-composition, the prepared FeSe$_{0.5}$Te$_{0.5}$ powder was mixed with 5 wt% ($x$ = 0.05), 10 wt% ($x$ = 0.1), 20 wt% ($x$ = 0.2), 30 wt% ($x$ = 0.3), and 40 wt% ($x$ = 0.4) Pb (99% purity of Pb powder). The mixed powder of Pb/FeSe$_{0.5}$Te$_{0.5}$ was pressed into cylinder pellets under 6000 Kg of pressure, which were sealed again into the evacuated quartz tube. Further sintering of these pellets was performed at the same temperature of 600°C for 4 h followed by furnace cooling to room temperature. The final pellets had a diameter of ~12 mm with a thickness of ~2.5 mm. All chemical manipulations were performed into an argon-filled glove box. This process was used to prepare several samples from different batches, all of which demonstrated high consistency in terms of superconducting properties.

The obtained samples were measured by the powder X-ray diffraction method (XRD) and carried out on the Rigaku SmartLab 3kW diffractometer filtered Cu-Kα radiation (wavelength: 1.5418 Å) and a Dtex250 linear detector were used. The profile analysis, the quantitative values of impurity phases (%), and the lattice parameter analysis of the measurement results were performed using Rigaku's PDXL software and the ICDD PDF4+ 2021 standard diffraction patterns database, and also from Rietveld refinements on the XRD data employing the Fullprof software suite [30]. Microstructural characterizations were performed using a field-emission scanning electron microscope. Magnetic measurements up to 9 T were performed by Quantum Design PPMS using Vibrating Sample Magnetometer (VSM) in the temperature range of 5-22 K under zero-field and field-cooling conditions. The zero-field cooled (ZFC) magnetization was measured by cooling the bulk sample in a zero field to 5 K, then applying a magnetic field and collecting data during the warming up process. The temperature dependence of the resistivity in a zero magnetic field was measured in a closed-cycle refrigerator in the temperature range of 7 to 300 K with a zero magnetic field. The electrical contacts were made using thin copper wires with silver epoxy. For resistivity and magnetic measurements, we had cut the sample into a rectangular shape. The typical dimensions of the samples for the resistivity measurements were as follows: 10.9 mm × 3.3 mm × 2.4 mm, the distance between electric contacts = 2.7 mm for $x = 0$; 6 mm × 3.4 mm × 2.5 mm, the distance between electric contacts = 1 mm for $x = 0.05$; 10.8 mm × 4.8 mm × 2.4 mm, the distance between electric contacts = 2.2 mm for $x = 0.1$; 10.3 mm × 3.4 mm × 2.4 mm, the distance between electric contacts = 2.8 mm for $x = 0.2$; 11 mm × 3.4 mm × 2.5 mm, the distance between electric contacts = 2.8 mm for $x = 0.3$; 10.9 mm × 3.0 mm × 1.9 mm, the distance between electric contacts = 3.2 mm for $x = 0.4$ .

## Results and discussion

Powder X-ray diffraction patterns were recorded for each prepared sample $FeSe_{0.5}Te_{0.5}$ + $x$Pb which are shown in Figure 1(a). The parent compound $FeSe_{0.5}Te_{0.5}$ ($x = 0$) exhibits an almost pure tetragonal phase with a very small amount (~4%) of the hexagonal phase, which can hardly be eliminated as previously reported [11, 27, 16]. The measured and refined XRD patterns of the parent compound ($x = 0$) and the sample with $x = 0.05$ are shown in Figure 1(b)-(c). Apart from the main tetragonal phase presented in all samples, a tiny amount of hexagonal phase was observed in the parent compound and the sample with $x = 0.05$ had a very small amount (~4%) of the PbTe type of cubic phase, in fact probably a solid solution Pb(Te$_{1-x}$,Se$_x$). The obtained lattice parameters of FeSe$_{0.5}$Te$_{0.5}$ polycrystalline ($a = 3.7950$ Å, $c = 5.9713$ Å) are nearly identical to those reported for FeSe$_{0.5}$Te$_{0.5}$ single crystals ($a = 3.815$ Å, $c = 6.069$ Å) and polycrystalline ($a = 3.7909$ Å, $c = 5.9571$ Å) [11]. However, the small amount of Pb-addition ($x = 0.05$) surprisingly removes the hexagonal phase completely, and it's not observed even at higher Pb-additions. It suggests that Pb works as a promotor for the superconducting tetragonal phase similar to the behaviour of other metal added iron-based superconductors [27, 28, 31, 32, 19]. The list of other presented phases is mentioned in table 1. In higher Pb addition, three extra phases such as PbTe, FeSe$_{1-\delta}$, and Fe appeared as the impurity phases. All patterns were fitted well with a tetragonal PbO structure type superconducting phase. The lattice parameters of the parent compound FeSe$_{0.5}$Te$_{0.5}$ are higher than those of the FeSe phase ($a = 3.7696$ Å, $c = 5.5201$ Å) [11], indicating Te substitution at Se-sites. A comparative analysis of the XRD patterns of the parent compound and Pb-added compounds, as shown in Figure 1(a), suggests that there is no deviation in the diffracted peaks due to Pb additions. On this basis, we can say that Pb does not enter the crystal lattice of the tetragonal phase of FeSe$_{0.5}$Te$_{0.5}$. In Pb added samples, however, PbTe is a dominant impurity phase, implying a slightly lower Te content in the FeSe$_{0.5}$Te$_{0.5}$ composition (table 1). These extra Te were observed in the form of PbTe, which is enhanced with Pb-additions. Besides, small amounts of FeSe$_{1-\delta}$ and Fe are also observed for $x = 0.3$ and $0.4$ samples, which are increased with Pb-additions. Figure 1(d) shows the obtained lattice parameters for multiple samples, confirming a very small divergence in lattice parameters with Pb additions. It could be owing to the parent compound FeSe$_{0.5}$Te$_{0.5}$'s somewhat lower Te/Se/Fe content. One should keep in mind that with larger Pb additions, the refinement error is slightly higher due to the presence of multiple impurity phases. It's also worth noting that Pb-addition promotes the creation of a tetragonal phase similar to Sn-addition [27] while simultaneously lowering Fe/Te/Se concentrations at high Pb-additions. In table 1, we also mentioned the crystallite size of the superconducting phase obtained from the XRD fitting data. The sample with $x = 0.05$ had a larger crystal size, but the crystallite size was decreased with further Pb additions.

Elemental analysis of a polycrystalline sample using energy-dispersive X-ray spectroscopy (EDX) offers the opportunity to measure the stoichiometric composition of the elements. For each sample, the EDX and the area of mapping were measured. The elemental mapping for the parent compound with $x = 0$ and the Pb-added sample with $x = 0.05$ and $0.1$ is shown in Figures 2 (i)-(ii), respectively. These figures confirm the homogeneous distribution of the constituent elements for $x = 0$ and $0.1$. However, the sample with $x = 0.05$ has a non-homogeneous distribution of Pb due to a very small amount of Pb-addition and very few areas were observed with Pb and Te rich, as shown in Figure 2(ii), which suggests the PbTe impurity phase as obtained from XRD measurements. For $x = 0.1$ and $0.2$, the distribution of Pb is almost homogeneous. At higher Pb concentrations ($x > 0.2$), however, a slightly different Fe/Te/Se ratio is observed, as well as a non-homogeneous distribution of the constituent elements, which supports the enhancement of the PbTe/Fe/FeSe$_{1-\delta}$ phase similar to XRD data analysis. From this data, the molar ratio Fe:Se:Te is found to be 1:0.49:0.51 for $x = 0$, which is slightly deviated for Pb-added samples.

Images of backscattered scanning electron microscopy (BSE-SEM, revealing chemical contrast) of the various polycrystalline FeSe$_{0.5}$Te$_{0.5}$ samples with various Pb additions have been performed. Microstructural analysis was performed on well-polished surfaces by using micron sandpaper without lubricants. Backscattered electron (BSE) images of FeSe$_{0.5}$Te$_{0.5}$ ($x = 0$) and FeSe$_{0.5}$Te$_{0.5}$ + $x$Pb from low to high magnification are shown in Figures 3(a)–(c) for $x = 0$, 3(d)-(f) for $x = 0.05$ and 3(g)–(i) for $x = 0.1$, respectively. These images contain light gray, white, and black contrasts corresponding to FeSeTe, PbTe, and pores, respectively. It appears that the microstructure is fairly homogeneous on the micro-scale for $x = 0$, and $x = 0.05$, where the light gray and black contrasts were observed only. As shown in Figure 3(a)–(c), the parent compound has many well-connected and disk-shaped grains with an average grain size of 2–4 μm, and pores do exist in some places. A small amount of Pb-addition ($x = 0.05$) improves grain connectivity and density by reducing pores, but the grain size is nearly the same as the parent compound, as depicted in Figure 3(d)-(f). Figures 3(g)-(i) show one of the most prominent phases of PbTe in the sample with $x = 0.1$ as a white contrast and many pores do exist as a black contrast. These impurity phases appeared at random, at grain boundaries, and within grains. The pores and impurity phases lead to poorer connections between the corresponding grains, as in the case of $x = 0.1$. The grains are plate-shaped, with an average grain size of ~1-2 μm. With a further increase in Pb content ($x > 0.1$), the grain size becomes smaller ~1 μm and the white contrast (PbTe) is observed in many places, with a larger area in between grains. Generally, these enhanced impurities lying between $FeSe_{0.5}Te_{0.5}$ grains significantly reduce grain-to-grain connections and give rise to a pronounced resistance to intergranular supercurrent paths. However, we have not observed any micro-cracks between the grains, as it is well known from other iron-based superconductors that extensive cracking occurs sometimes at grain boundaries and sometimes within grains [33, 34]. BSE images revealed that a small amount of Pb addition improves grain connectivity and reduces pores, whereas a larger amount of Pb addition decreases phase purity and cleanness of grain boundaries and reveals many pores. There are non-superconducting obstructions at the grain boundaries of $FeSe_{0.5}Te_{0.5}$ at higher Pb additions, similar to the case of Pb-added $Sr_{0.6}K_{0.4}Fe_2As_2$ (Sr122) [35]. Higher Pb addition ($x > 0.05$), as shown in Figure 3, increased the amounts of impurity phases, posing a problem in improving grain connectivity. These results are similar to those discussed above with XRD data. Assuming the theoretical density of $FeSe_{0.5}Te_{0.5}$ is 6.99 g/cm$^3$ [7], the calculated densities of the prepared $FeSe_{0.5}Te_{0.5}$ + $x$Pb polycrystalline sample are around 51%, 58.4%, 60%, 61.7%, 70%, 72% for $x = 0$, 0.05, 0.1, 0.2, 0.3, and 0.4, respectively. One should note that these densities are calculated by assuming the pure phase of $FeSe_{0.5}Te_{0.5}$ only. The sample density is slightly improved with increasing Pb content, and the area of the impurity phase is also enhanced. Based on the foregoing, we can conclude that a very small amount of adding Pb to the $FeSe_{0.5}Te_{0.5}$ sample reduces the wetting phase at grain boundaries and micro/nanopores, resulting in a slightly increased sample density and improved grain size and grain connectivity.

Figure 4(a)–(c) depicts the zero-field resistivity as a function of temperature for nominal compositions of $FeSe_{0.5}Te_{0.5}$ + $x$Pb ($x = 0$–0.4). The parent $FeSe_{0.5}Te_{0.5}$ ($x = 0$) showed a broad anomaly in the resistivity at a temperature of around 110 K owing to the structural phase transition [17]. At the same time, Pb addition is gradually changing the electrical behaviour of the samples, and this anomaly also seems to be evident in the 5% Pb-added samples ($x = 0.05$), and finally disappears in 10wt% Pb-added ($x = 0.1$) and higher Pb-added samples, which could be due to the presence of impurity phases. A slightly higher value of the normal state resistivity (Figure 4(a)) for $x = 0.05$ may be possible due to the presence of a small amount and inhomogeneous distribution of PbTe phase. As further Pb additions are increased, such as $x = 0.1$, the amount of PbTe phase is enhanced and its distribution inside the sample becomes more homogeneous (as clear from the microstructural analysis), which slightly improves the whole sample density, so it could be a reason to decrease the normal state resistivity. Besides, the resistivity of the higher Pb-added samples is much lower than that of the Pb-free sample within the whole measured temperature range, while the PbTe phase has a semiconductor trend. The addition of Pb increases the density of samples, as discussed in the microstructural analysis, which lowers the resistivity of the sample and supports the content of the superconducting tetragonal phase. The positive slope of resistivity in the higher Pb-added sample indicates that it mainly exhibits metallic behaviour throughout the normal state. The reduced normal state resistivity behaviour was also reported for Sn added $FeSe_{0.5}Te_{0.5}$ bulk samples [27].

Figure 4(b) depicts the detailed resistivity as a function of temperature ranging from 5 K to 22 K. The corresponding $T_c^{onset}$ and $T_c^{offset}$ values for various samples are shown in table 2. It is obvious that all the samples exhibit a superconducting transition. The overall residual resistivity ratio (RRR) is slightly reduced with Pb-additions, as similar to the reported Ag addition [36]. Residual resistivity ($\rho_0$) is a significant material parameter that is obtained, here, by extending the normal state resistivity behaviour up to absolute zero temperature. Interestingly, the residual resistivity ($\rho_0$) value is around 0.3 mΩ-cm for $x = 0$ and 0.7 mΩ-cm for $x = 0.05$, but with higher Pb additions, its value increased ($\rho_0 = 0.9$ mΩ-cm, $x = 0.1$; $\rho_0 = 1.1$ mΩ-cm, $x = 0.2$; $\rho_0 = 1.6$ mΩ-cm, $x = 0.3$; $\rho_0 = 2.3$ mΩ-cm, $x = 0.4$;) which might be due to impurity and imperfection scattering at zero temperature. It is observed that Pb-free and 5% Pb-added samples exhibit comparable values of the $T_c^{onset}$ which is a little lower than that of the Pb-free sample. What is more notable is that their $T_c^{offset}$ values are quite different. The Pb added samples exhibit a relatively broad transition with a low $T_c^{offset}$. Meanwhile, 5 wt% Pb-added sample shows a 1 K lower transition but shows a comparatively sharper transition with a $T_c$ of 13.8 K and a $T_c^{offset}$ of 11.7 K (table 2). The transition width of 2.1 K suggests a sharper transition than for the pure sample. The sharper transition in the 5wt% Pb added sample suggests that this sample ($x = 0.05$) has better grain connections than the Pb-free one ($x = 0$). On the other hand, the increment of Pb addition exhibits the broadening of superconducting transition with no zero resistance observed in the measured temperature as low as 8 K, which might result due to the increased impurity phase (PbTe) and the decreased superconducting phase. The decrease in the transition temperature $T_c$ could be possible due to the reduced Te concentration in the $FeSe_{0.5}Te_{0.5}$ composition, as observed from the slightly reduced lattice parameters, as depicted in Figure 1(b). The previous study on the Li doping effect on $FeSe_{0.5}Te_{0.5}$ has shown that Li can enter the crystal structure [26]. As a result, the onset transition of the superconducting transition can be raised by about 1-1.5 K by 1 wt% Li doping, whereas there is no enhancement on $T_c^{offset}$. The 3d metals, such as Co and Ni, even suppress the superconductivity of FeSe$_{0.5}$Te$_{0.5}$ [11]. Compared to these previous studies, 5% Sn addition can dramatically increase $T_c^{offset}$ of FeSe$_{0.5}$Te$_{0.5}$ by 3 K and also enhance $T_c^{on}$ without changing the crystal structure of FeSe$_{0.5}$Te$_{0.5}$ which implies that Sn seems to be the most promising additive among metals to further improve the superconductivity in the 11-type IBS [27] and also for 1111 (REFeAsO; RE = rare earth) [32].

To better understand the intergrain connections in these polycrystalline samples, we show the temperature dependence of resistivity with respect to various currents ($I$ = 5, 10, and 20 mA) in Figure 4(c). Generally, $T_c^{onset}$ represents the particular grain effect (intragrain effect) and $T_c^{offset}$ is related to the grain connections (intergrain effect) [32, 37]. The parent compound has almost no broadening with the current, whereas the sample $x$ = 0.05 has almost the same order of broadening compared to the parent compounds but has a sharper transition. It suggests that low amounts of Pb added to samples ($x$ = 0.05) have better grain connectivity compared to the parent compound ($x$ = 0), as clearly observed from the microstructural analysis. Further Pb additions enhanced this broadening, as shown in Figure 5(c) for $x$ = 0.1 and 0.2. Interestingly, the $T_c^{offset}$ is more sensitive to the applied current at higher Pb additions ($x$), implying that Pb additions ($x$ > 0.05) reduce the grain connections due to the enhancement of impurity phases and pores, as evidenced by microstructure and XRD studies.

Figures 5(a) shows the zero-field cooling (ZFC) and field-cooling (FC) magnetization curves for two samples, $x$ = 0 and $x$ = 0.05, measured under an applied magnetic field of 20 Oe. The ZFC magnetization curve indicates a superconducting transition at 14 K for $x$ = 0, but at 13.3 K for $x$ = 0.05, which is slightly wider than that of $x$ = 0. This single-step transition can be interpreted in terms of the intergranular properties of these samples, as discussed in detail elsewhere [38]. With the addition of Pb, the $T_c$ decreased, which could be because of changes in the concentration of Te [17, 11] as discussed in the XRD and resistivity sections.

To calculate the critical current density $J_c$, the hysteresis loop at 7 K for $x$ = 0 and 0.05 is measured with the rectangular-shaped sample, as depicted in the inset of Figure 5(a). These loops display a combination of ferromagnetic and superconducting signals, which is consistent with the reported results of FeSe family [16, 39]. The width of the hysteresis loop $\Delta m(H)$ is small, which could be due to either poor intergrain connectivity or weak intragranular pinning. We calculated the critical current density $J_c$ for our samples on the basis of $J_c$ = 20$\Delta m$/$Va$(1 − $a$/3$b$) [40], where $a$ and $b$ are the lengths of the shorter and longer edges, respectively, $V$ is the volume of the sample, and $\Delta m$ is the hysteresis loop width. In Figure 5(b), the magnetic field dependence of the critical current density ($J_c$) at 7 K for two samples ($x = 0$ and 0.05) is compared up to 9 T. Note that Pb added FeSe$_{0.5}$Te$_{0.5}$ ($x = 0.05$) enhanced the $J_c$ value across the entire field range up to 9 T and shows the field dependence that is nearly identical to the parent compound. The estimated $J_c$ value is ~8.51 A/cm$^2$ and ~30.1 A/cm$^2$ for $x = 0$ and 0.05, respectively, at $T = 7$ K and $H = 2$ T. The enhanced $J_c$ must mean the improved behaviour of the flux pinning. We have calculated the field dependency of the vortex pinning force density, $F_p$, given by $F_p = \mu_0 H \times J_c$ [41] at 7 K which is shown in the inset of Figure 5(b). It demonstrates that a small amount of Pb added samples enhanced the pinning force in the entire field region when compared to that of the parent compounds ($x = 0$), which agrees nicely with the $J_c$ enhancement results. Moreover, comparing the $F_p$ values of the parent compound with those reported in the literature [42, 39], it can be noted that they are similar to those reported for bulk and polycrystalline Fe(Se,Te) samples (0.1-1 GN/m$^3$). Furthermore, this improvement in $J_c$ and $F_p$ can also be related to the increased density and improved grain connections caused by the addition of a small amount of Pb, which is clearly observed in the microstructural analysis and resistivity studies. These results are similar to the enhancement of the reported $J_c$ values for Sn added SmFeAs(O,F) [32] bulk samples. Further $J_c$ enhancement of these Pb-added FeSe$_{0.5}$Te$_{0.5}$ samples could be possible via the optimization of heating patterns, the densification by high-pressure synthesis, as well as sintering under ambient pressure or hot isostatic pressing (HIP), and the formation of a c-axis grain aligned microstructure [4].

## Conclusion

In summary, the superconducting properties of FeSe$_{0.5}$Te$_{0.5}$ have been studied under the influence of Pb metal additions by structural, electrical, and magnetic characterizations. Our results indicate that the superconducting transition is decreased and the impurity phase is enhanced with Pb addition due to the reduced Fe/Se/Te ratio from the stoichiometric FeSe$_{0.5}$Te$_{0.5}$ composition. A microstructural investigation and resistivity studies revealed that superconducting grains have a disc shape and the grain connectivity is reduced with high Pb additions compared to the parent compound. However, the low amount of Pb addition has less effect on the superconducting transition and the improved $J_c$ in the measured magnetic field (up to 9 T) due to the improved grain connections. Pb addition, in conclusion, reduces the superconducting properties of FeSe$_{0.5}$Te$_{0.5}$, but a small amount of Pb is effective in improving the intergranular behaviour and the critical current properties. Further work needs to be carried out on the optimization of the amount of very low Pb addition ($x < 0.05$) and processing parameters with the aim of further improving critical current properties.

**Acknowledgments:**

The work was supported by the National Centre for Research and Development (NCBR), Poland through Project No. POIR.04.01.02-00-0047/17 and by National Science Centre (NCN), Poland, grant number "2021/42/E/ST5/00262" (SONATA-BIS 11). S.J.S. acknowledges financial support from National Science Centre (NCN), Poland through research Project number: 2021/42/E/ST5/00262.

**Table 1:**

List of the impurity phases and crystallite size of the main phase in FeSe$_{0.5}$Te$_{0.5}$ + $x$Pb samples. The quantitative values of impurity phases (%) and crystallite size of the measurement results are performed using Rigaku's PDXL software and the ICDD PDF4+ 2021 standard diffraction patterns database.

| Sample | Hexagonal (H) (%) | PbTe (%) | Fe (%) | FeSe$_{1-\delta}$ (%) | Crystallite Size (nm) (FeSe$_{0.5}$Te$_{0.5}$ phase) |
|---|---|---|---|---|---|
| $x = 0$ | 3-4 | - | - | - | 34 |
| $x = 0.05$ | - | 4-5 | - | - | 45 |
| $x = 0.1$ | - | 14.3 | - | - | 29 |
| $x = 0.2$ | - | 19 | 1.7 | - | 30 |
| $x = 0.3$ | - | 34 | 3.9 | 2.3 | 28 |
| $x = 0.4$ | - | 42 | 6.3 | 3.9 | 23 |

**Table 2:**

List of room temperature resistivity ($\rho_{300K}$), resistivity at 20 K ($\rho_{20K}$), Residual Resistivity Ratio ($RRR = \rho_{300K}/\rho_{20K}$), the onset transition temperature ($T_c^{onset}$), the offset of transition temperature ($T_c^{offset}$) and the transition width ($\Delta T$) for FeSe$_{0.5}$Te$_{0.5}$ + $x$Pb.

| Sample FeSe$_{0.5}$Te$_{0.5}$ + $x$Pb | $\rho_{300K}$ (m$\Omega$)-cm | $\rho_{20K}$ (m$\Omega$)-cm | $RRR$ ($\rho_{300K}/\rho_{20K}$) | $T_c^{onset}$ (K) | $T_c^{offset}$ (K) | $\Delta T$ (K) |
|---|---|---|---|---|---|---|
| $x = 0$ | 1.8 | 0.9 | 2.0 | 15 | 11.9 | 3.1 |
| $x = 0.05$ | 2.0 | 1.1 | 1.8 | 13.8 | 11.7 | 2.1 |
| $x = 0.1$ | 1.8 | 1.9 | 0.9 | 12.6 | 8.6 | 4 |
| $x = 0.2$ | 1.5 | 1.3 | 1.1 | 11.73 | <8.6 | >3.1 |
| $x = 0.3$ | 1.2 | 0.8 | 1.6 | 10.5 | <8.6 | >3.1 |
| $x = 0.4$ | 0.5 | 0.2 | 2.1 | 10.3 | <8.6 | >3.1 |

**Figure: 1 (a)** Powder X-ray diffraction patterns (XRD) of FeSe$_{0.5}$Te$_{0.5}$ + $x$Pb ($x$ = 0, 0.05, 0.1, 0.2, 0.3 and 0.4). **(b)** Experimental and calculated diffraction patterns and their differences for the room temperature X-ray diffraction data for $x$ = 0. Fe$_{1.1}$Se$_{0.5}$Te$_{0.5}$ tetragonal phase was determined as the actual composition of the superconducting phase rather than the nominal composition of FeSe$_{0.5}$Te$_{0.5}$. A hexagonal phase of Fe$_{0.6}$Se$_{0.54}$Te$_{0.46}$ (~4-5%) has been also identified, while another hexagonal phase, Fe$_7$Se$_8$, was not included in the refinement due to very weak reflections. **(c)** Experimental and calculated diffraction patterns and their difference for the room temperature X-ray diffraction data for $x$ = 0.05. PbTe phase was also observed from the refinement which is around 4-5%, but any hexagonal phase was not observed. **(d)** The variation of lattice constants '$a$' and '$c$' of main tetragonal phase as function of Pb weight concentration ($x$). The red hexagonal data point for FeSe$_{0.5}$Te$_{0.5}$ is taken from Reference [11] to compare the lattice parameters with our data. The error in the lattice parameters is up to ~0.1 %.

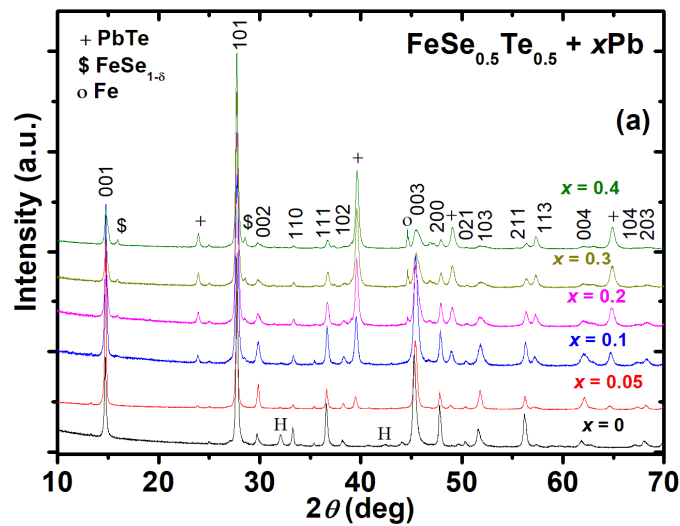

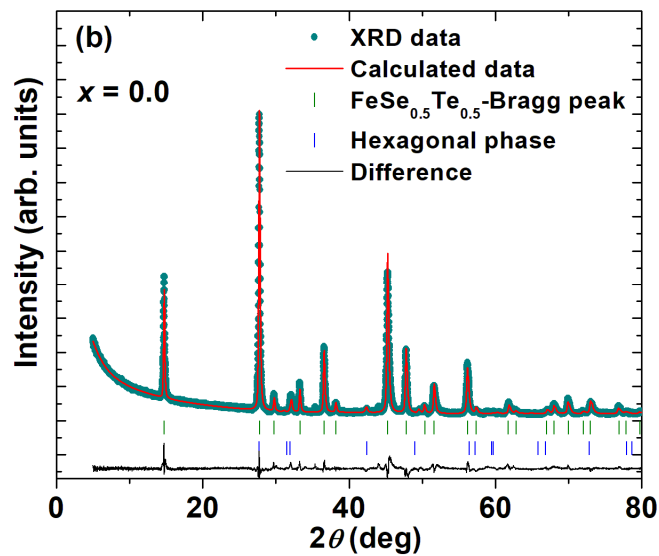

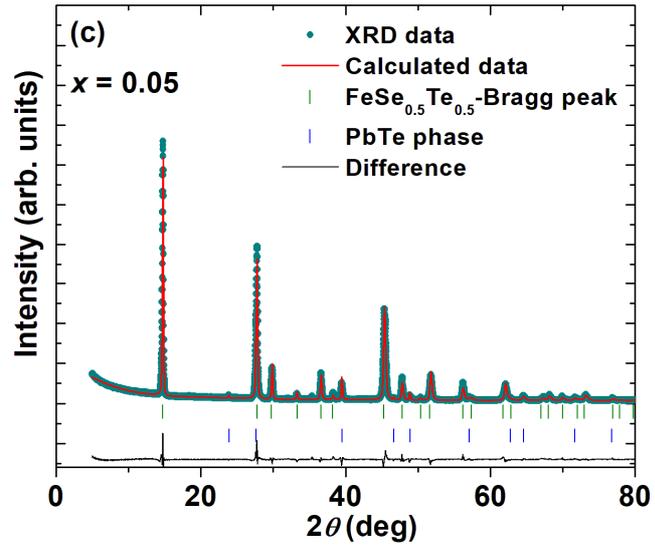

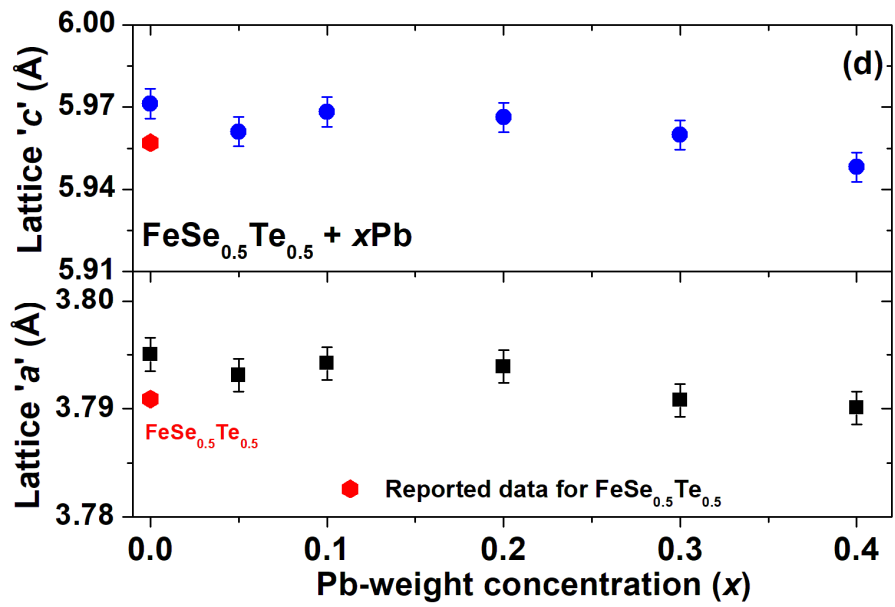

**Figure:2 (i)** Mapping for the constituent elements of FeSe$_{0.5}$Te$_{0.5}$ + $x$Pb ($x$ = 0).

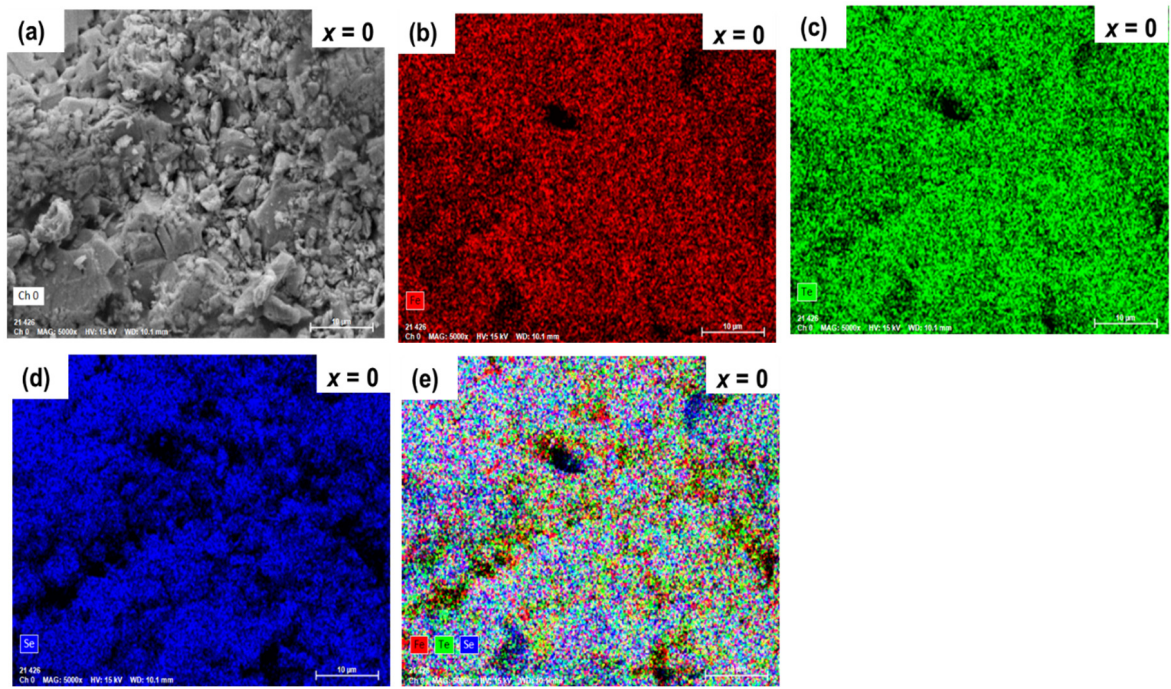

**(ii)** Mapping for the constituent elements of FeSe$_{0.5}$Te$_{0.5}$ + $x$Pb ($x$ = 0.05).

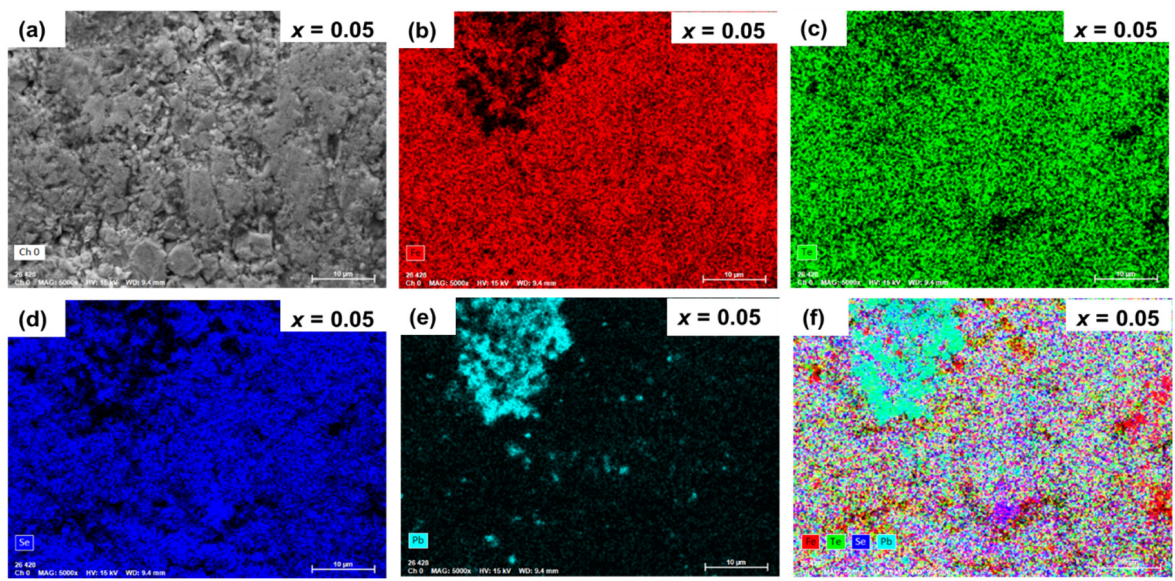

**(iii)** Mapping for the constituent elements of FeSe$_{0.5}$Te$_{0.5}$ + $x$Pb ($x = 0.1$).

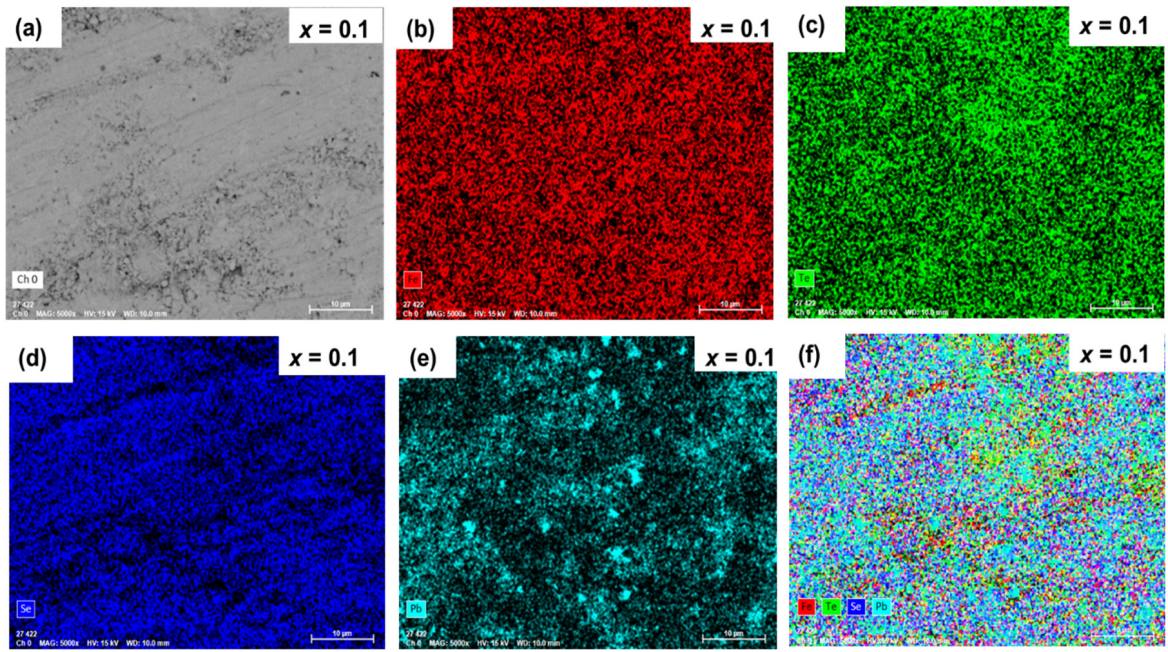

**Figure:3** Back-scattered (BSE) images of FeSe$_{0.5}$Te$_{0.5}$ + $x$Pb: (a)-(c) for $x = 0$; (d)-(f) for $x = 0.05$; (g)-(i) for $x = 0.1$.

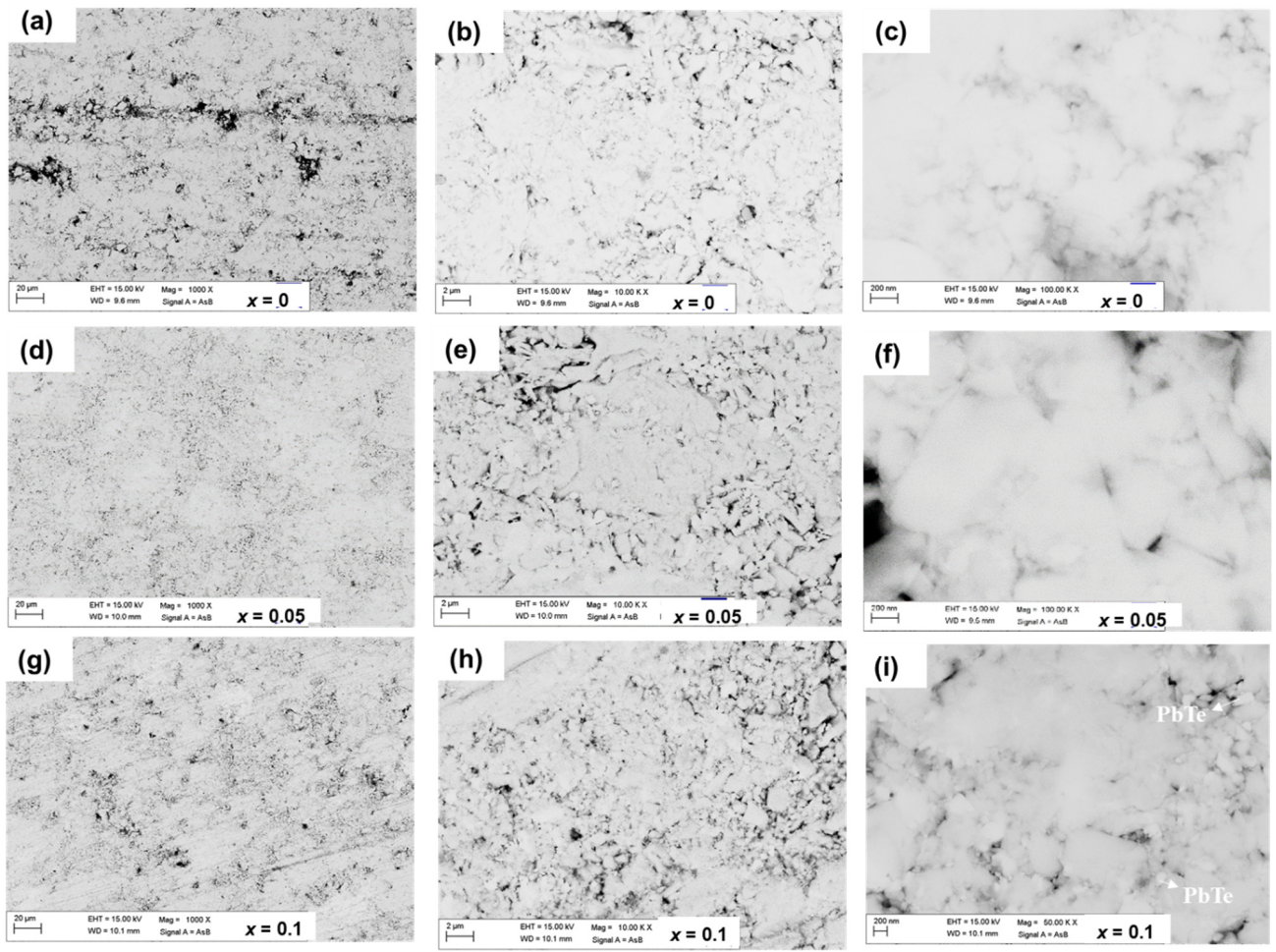

**Figure: 4 (a)** The temperature dependence of resistivity ($\rho$) up to room temperature for various FeSe$_{0.5}$Te$_{0.5}$ with Pb ($x = 0$-0.4). **(b)** Low temperature resistivity behaviours with temperature for various samples. **(c)** The temperature dependence of low temperature resistivity for various currents $I = 5, 10, 20$ mA for FeSe$_{0.5}$Te$_{0.5}$ + $x$ Pb ($x = 0, 0.05. 0.1$ and $0.2$).

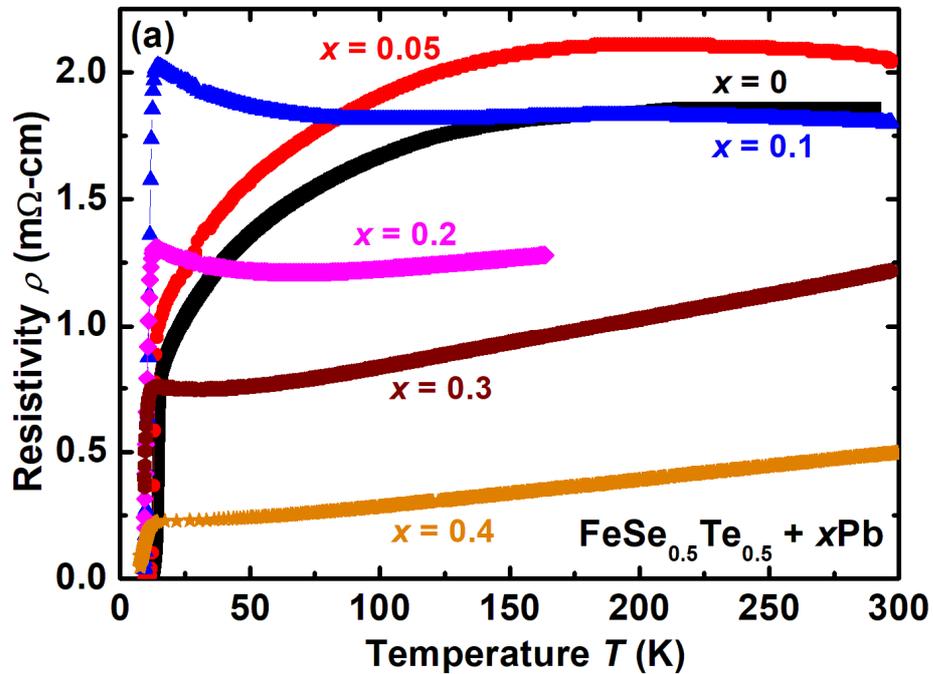

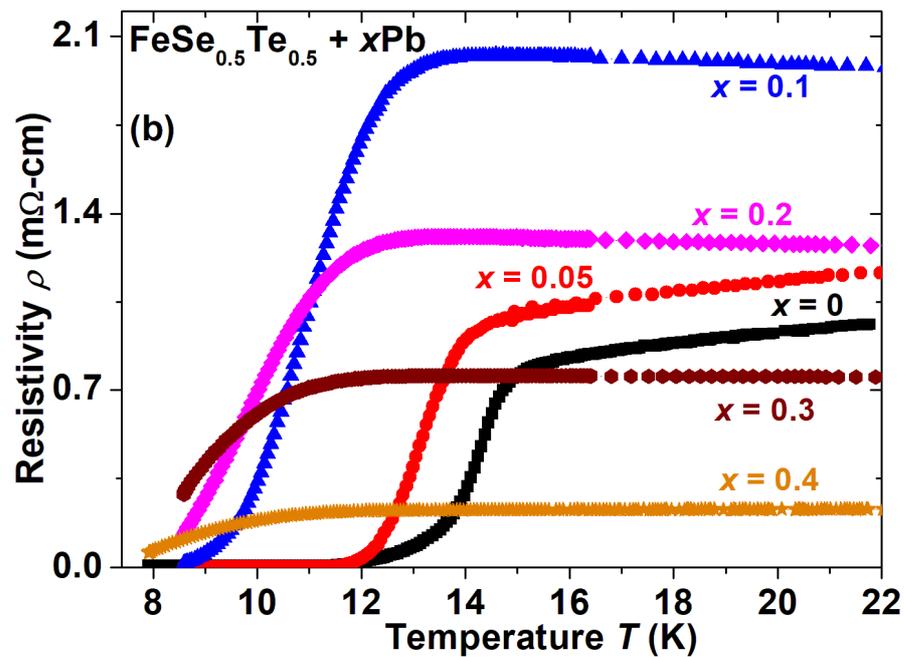

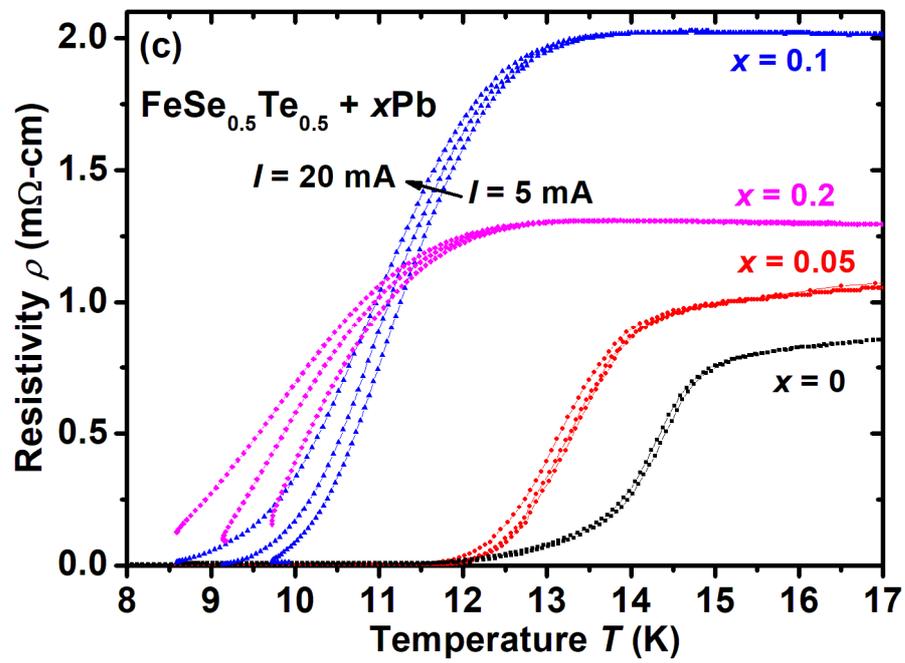

**Figure:5 (a)** Temperature dependence of the magnetic susceptibility of FeSe$_{0.5}$Te$_{0.5}$ ($x = 0$) and FeSe$_{0.5}$Te$_{0.5}$ + 5% weight Pb ($x = 0.05$) in a magnetic field of 20 Oe. The inset figure shows the magnetic hysteresis loop $M(H)$ at 7 K for $x = 0$ and $x = 0.05$. **(b)** The dependence of critical current density ($J_c$) with magnetic field ($H$) for parent FeSe$_{0.5}$Te$_{0.5}$ and 5% Pb weight samples ($x = 0.05$) at 7 K. The calculated pinning force density $F_p$ at 7 K as a function of magnetic field is depicted in the inset figure.

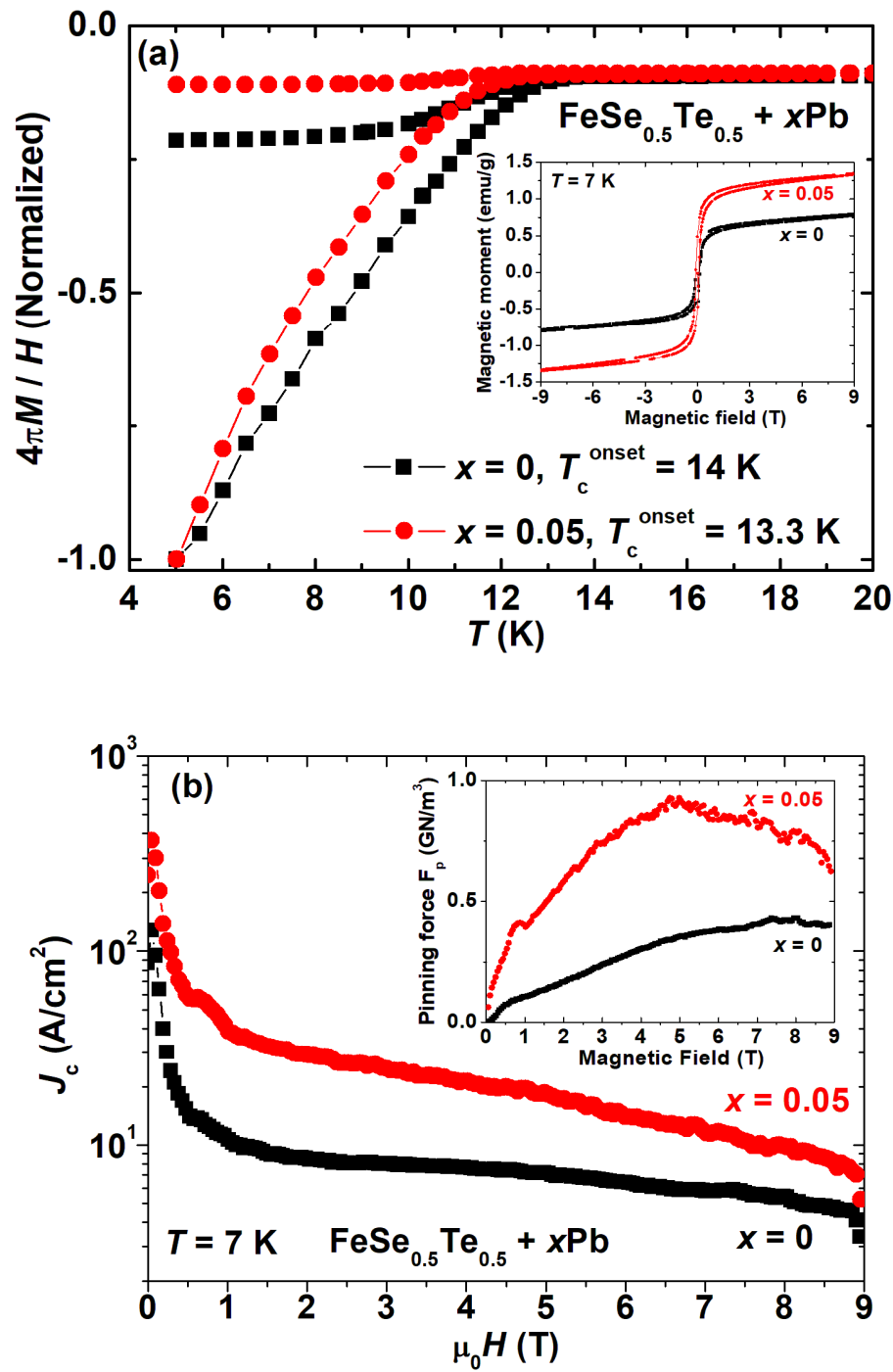